\documentclass[10pt,aip,prb,twocolumn, nofootinbib, reprint]{revtex4-1}

\usepackage{amssymb,amsmath}
\usepackage{subfig}
\usepackage{graphicx}
\usepackage{color}
\usepackage{hyperref}
\usepackage{listings}
\usepackage{multirow}
\usepackage{etoolbox}
\usepackage[english]{babel}

\usepackage{braket}
\usepackage{array}
\usepackage{booktabs}
\usepackage{threeparttable}
\usepackage{bm}
\usepackage{epstopdf}
\usepackage{textcomp}

\hypersetup{
    bookmarks=true,         
    unicode=false,          
    pdftoolbar=true,        
    pdfmenubar=true,        
    pdffitwindow=false,     
    pdfstartview={FitH},    
    colorlinks=false,       
    linkcolor=black,          
}
\graphicspath{{figures/}}

\makeatletter
\def\@email#1#2{%
 \endgroup
 \patchcmd{\titleblock@produce}
  {\frontmatter@RRAPformat}
  {\frontmatter@RRAPformat{\produce@RRAP{*#1\href{mailto:#2}{#2}}}\frontmatter@RRAPformat}
  {}{}
}%
\makeatother

\begin{document}

\definecolor{dkgreen}{rgb}{0,0.6,0}
\definecolor{gray}{rgb}{0.5,0.5,0.5}
\definecolor{mauve}{rgb}{0.58,0,0.82}

\lstset{frame=tb,
  	language=Matlab,
  	aboveskip=3mm,
  	belowskip=3mm,
  	showstringspaces=false,
  	columns=flexible,
  	basicstyle={\small\ttfamily},
  	numbers=none,
  	numberstyle=\tiny\color{gray},
 	keywordstyle=\color{blue},
	commentstyle=\color{dkgreen},
  	stringstyle=\color{mauve},
  	breaklines=true,
  	breakatwhitespace=true
  	tabsize=3
}

\title[Long-Lived Electronic Coherences from First Principles]{Long-Lived Electronic Coherences from First Principles}
\author{Jiří Suchan}
 \affiliation{Institute for Advanced Computational Science, Stony Brook University, Stony Brook, New York 11794, United States}
 \author{Benjamin G. Levine}
  \email{ben.levine@stonybrook.edu}
  \affiliation{Institute for Advanced Computational Science, Stony Brook University, Stony Brook, New York 11794, United States}
\affiliation{Department of Chemistry, Stony Brook University, Stony Brook, New York 11794, United States}

\date{\today}

\begin{abstract}
Electronic coherences can be leveraged to control molecular dynamics, but such control is limited by ultrafast decoherence driven by coupling between electronic excitations and molecular vibrations. With the goal of understanding and controlling electronic coherence in molecules, we introduce a first-principles approach that enables direct simulation of the creation and decay of electronic coherences in molecules.  Simulations of long-lived experimentally-observed coherences created upon multiphoton excitation of thiophene reveal coherent electronic motions within a dense manifold of Rydberg states, enabled by their relatively parallel potential energy surfaces.  

\end{abstract}

\maketitle

\section{Introduction}
Electronic coherences can be leveraged to control light-driven dynamics in both natural\cite{duan2017nature,maiuri2018coherent} and synthetic\cite{vacher_1,lepine2014attosecond} molecules. In an isolated atom, coherences can survive for hundreds of picoseconds or more.\cite{tenwolde1989} Yet in molecules with more than a few atoms, the loss of electronic coherence can be ultrafast, with the vibrations of the molecule acting as a bath that kills coherence.\cite{bittnerrosskydecoh1}  With an eye towards control, it would be extremely desirable to understand the factors determining the rate of electronic decoherence in excited molecules.

Using time-resolved photoelectron spectroscopy, Kaufman et al.\ recently observed long-lived electronic coherences in thiophene (C$_4$SH$_4$) following multiphoton excitation by a strong field laser pulse.\cite{kaufman2023longthio}  The pump pulse employed in these experiments generated coherences between two or more states separated by roughly one photon energy (1.55 eV), and the use of a phase-locked probe pulse confirmed their presence even 200 fs after the initial excitation. Such lifetimes appear exceptionally long, spanning many vibrational periods. 

Kaufman et al.\cite{kaufman2023longthio} tentatively suggest that either Rydberg states or valence states involving lone pair orbitals may be involved in the coherence. Simulation can fill these knowledge gaps by providing a more detailed picture of the dynamics. However, accurate simulation of coherence and decoherence requires a dynamical treatment of electron-nuclear motion on many coupled electronic states. With 21 vibrational modes and roughly 20 energetically accessible electronic states, exact quantum simulation is unfeasible.  

An efficient approximate approach would be an ideal alternative, but modeling long-lived coherences in dense manifolds of states remains a challenge for standard simulation methods. 
To better understand what is required, we will discuss the origin of electronic coherence and decoherence in molecules.  Let us consider a molecular wavefunction populating two electronic states in the Born-Huang expansion:
\begin{equation}
\begin{split}
\Psi(\bm{r}, \bm{R}, t) = \chi_1(\bm{R},t) \psi_1(\bm{r}; \bm{R}) +\\ 
\chi_2(\bm{R},t) \psi_2(\bm{r}; \bm{R}) \,.
\end{split}
\end{equation}
Here $\psi_i$ and $\chi_i$ are the respective electronic and vibrational wavefunctions associated with electronic state $i$. Nuclear positions are denoted as $\bm{R}$, electronic positions as $\bm{r}$, and time as $t$.
One can show the electronic coherence between states 1 and 2 to be
\begin{equation}
\begin{split}
    \rho_{12}(t) =
    & a_1(t)a^*_2(t) \times \\
    & \int |\chi'_1(\boldsymbol{R},t)||\chi'_2(\boldsymbol{R},t) | 
    e^{i(\theta_1(\boldsymbol{R},t)-\theta_2(\boldsymbol{R},t))t}
    d\boldsymbol{R} \times \\
    & e^{i(\omega_2-\omega_1)t}.
\end{split}
\end{equation}
Here, we have factorized the vibrational wave functions according to 
\begin{equation}
\begin{split}
    \chi_i(\boldsymbol{R},t) & = a_i(t) e^{-i\omega_it} \chi'_{i}(\boldsymbol{R},t) \\
    & = a_i(t) e^{-i\omega_it} | \chi'_{i}(\boldsymbol{R},t) | e^{i\theta_i(\boldsymbol{R},t)t}.
\end{split}
\end{equation}
The coefficient of electronic state $i$ is represented $a_i(t)$, $\omega_i$ is the energy of the minimum of the potential energy surface (PES) of state $i$, $\chi'_{i}(\boldsymbol{R},t)$ is normalized, and $\theta_i(\boldsymbol{R},t)$ is the position-dependent phase angle of $\chi'_{i}(\boldsymbol{R},t)$:
\begin{equation}
    \theta_i(\boldsymbol{R},t)=\arg(\chi'_i(\boldsymbol{R},t)).
\end{equation}
This factorization allows us to clearly see three different mechanisms by which the coherence, $\rho_{12}(t)$, may decay to zero:\cite{vacher2017electron_vMCG,coh_linesderiv}  a) The amplitude, $a_i$, of one or both of the states may go to zero when population transfers to a different state; b) the $\boldsymbol{R}$-dependent phase factor, $e^{i(\theta_1(\boldsymbol{R},t)-\theta_2(\boldsymbol{R},t))t}$, may become rapidly oscillatory, in which case the integral over $\boldsymbol{R}$ becomes zero; or c)  The spatial overlap, $|\chi'_1(\boldsymbol{R},t)||\chi'_2(\boldsymbol{R},t) |$, may decay to zero because the wave packets are evolving on nonparallel PESs.  If these three mechanisms are slow enough, the coherence may be observed to oscillate according to a phase factor determined by the energy difference between electronic states, $\omega_2-\omega_1$, as well as the dynamics of the vibrational overlap, $\chi'_{1}(\boldsymbol{R},t)\chi'^*_{2}(\boldsymbol{R},t)$.

One can clearly see that a description of the vibrational wave packets on multiple electronic states is central to a correct quantitative description of electronic coherence and decoherence. Thus methods based on a fully quantum treatment of nuclear dynamics are a natural fit, providing an accurate description of coherence, albeit at a high computational cost.  In fact, exact quantum simulations of model systems have shed light on numerous general features of quantum decoherence in molecules, and also provide useful benchmarks for more approximate methods.\cite{Gu2018,Hu2018}  When the goal is to simulate the dynamics of a particular molecule, multi-configuration time-dependent Hartree (MCTDH)\cite{csehi2020preservation_MCTDH,dey2022quantum,arnold2017electronic} can provide an extremely accurate description of the dynamics of vibrational wave packets on multiple electronic states, but requires a fitted PES, often in reduced dimensionality to keep computational costs down.  Methods based on Gaussian wave packets\cite{vacher2017electron_vMCG,jenkins2018ehrenfest,Tempelaar2018,tran2020quantum,scheidegger2022search} provide a lower cost description in full dimensionality.  But they come with their own complexities, owing to the need for multiple trajectory basis functions evolving on different PESs to interact with one another.  Classical Ehrenfest simulations can provide useful insights into coherences on short timescales,\cite{jenkins2016charge, franco_oligo, Gu2017} but Ehrenfest does not incorporate decoherence and, therefore, cannot discern between long- and short-lived coherences.  This leads to well-known qualitative errors on longer timescales.\cite{tully1990molecular,miller2013}  Incorporating decoherence effects into Ehrenfest effectively solves the problem for two-state systems,\cite{zhu2004coherent,bedard2005mean} but properly treating long-lived coherences in systems with many electronic states is not straightforward.\cite{esch2020a} 
Thus, an on-the-fly classical trajectory-based approach capable of accurately describing both long-lived coherences and ultrafast decoherence has been elusive.

To fill this gap, we have introduced the Ehrenfest with collapse-to-a-block (TAB) nonadiabatic molecular dynamics approach,\cite{esch2020a,tab21} and in this work we apply it to simulate the long-lived coherence observed in excited thiophene. The TAB method has recently been combined with an {\em ab initio} treatment of the electronic structure for the first time.\cite{cycbut_ours} In the present work, we demonstrate TAB to be a computationally lightweight alternative to fully quantum methods, capable of discerning between long- and short-lived coherences. {\em Ab initio} TAB simulates the system on the fly in its full dimensionality and does not require fitted PESs. Used in conjunction with time-dependent configuration interaction (TDCI)\cite{tdci,durden2022} calculations of multiphoton excitation by the pump pulse, we are able to explicitly simulate the dynamics observed in the Kaufman experiment. With detailed simulations of coupled electron-nuclear dynamics in hand, we definitively assign the observed coherences to specific states and gain understanding of the physics underlying its long lifetime. 

\section{Methods and computational details}
\subsection{Nonadiabatic dynamics: TAB method}
Here, we highlight the key features of the method relevant to the present results.
The TAB method is based on the Ehrenfest molecular dynamics method.\cite{miller1978classical,meyera1979classical,micha1983self,isborn2007time} {Ehrenfest is an independent trajectory method where the 
nuclei are treated classically, and the equation of motion is
\begin{equation}
\label{eq:eom}
\frac{d}{dt}\left[ \sum_\mu \frac{P_\mu^2}{2m_\mu} + \bra{\psi(t)}\hat{H}_{\mathrm{e}}(\bm{R}(t))\ket{\psi(t)}_{\bm{r}} \right]=0.
\end{equation}
Here $P_\mu$ is the $\mu$th element of the classical nuclear momentum, $m_\mu$ is the associated mass, and $\hat{H}_{\mathrm{e}}$ is the electronic Hamiltonian.  The classical nuclear coordinates are denoted $\bm{R}(t)$, and the subscript $\bm{r}$ indicates integration over on the electronic coordinates.}
The electronic wave function is then evolved by solving the time-dependent electronic Schr\"{o}dinger equation:
\begin{equation}
i\hbar\frac{\partial \ket{\psi(t)}}{\partial t} = \hat{H}_{\mathrm{e}}(\bm{R}(t))  \ket{\psi(t)}\,.
\end{equation}
In the Ehrenfest method, there is no consideration of decoherence due to the loss of nuclear overlap (mechanism c) above).\cite{ehrenfest_chapter} Instead, trajectories evolve along the mean-field PES, which can result in unphysical outcomes.\cite{tully1990molecular,miller2013} We remedy these issues by incorporating the collapse-to-a-block (TAB) decoherence correction scheme.\cite{esch2020a,tab21,tabgauss} 

The specific version of the TAB collapse algorithm used here is described in detail ref \citenum{tabgauss}, where this correction is referred to as TAB-w1.  The TAB correction assumes Gaussian decay of pairwise electronic coherences with a rate governed by the difference of the gradients of associated PESs, as originally derived by Rossky and Bittner:
\begin{equation}
\label{eq:decohtime}
\tau_{ij}^{-2}=\sum_{\eta}\frac{\left(F_{i,\eta}-F_{j,\eta}\right)^2}{8\hbar^2\alpha_{\eta}}\,,
\end{equation}
where $\eta$ indexes vibrational degrees of freedom, $F_{i,\eta}$ is the corresponding force matrix element on state $i$, and $\alpha_{\eta}$ is a parameter that represents the width of a Gaussian vibrational wave packet. 
This expression has a very intuitive interpretation; the more different the slopes of the PESs of two states, the faster their vibrational overlap, $\chi'_{1}(\boldsymbol{R},t)\chi'^*_{2}(\boldsymbol{R},t)$, will decay.  This and similar approximations have been used for years to correct the nonadiabatic transition probabilities incorporated into simulations of nonadiabatic molecular dynamics.\cite{zhu2004coherent,bedard2005mean,ouyang2015surface,min2015coupled,subotnik2016understanding,wang2016recent,martens2019hopping,vindel2021decoherence}  The novel feature of TAB is that each state pair is assigned an independent decoherence time, allowing for an accurate treatment of decoherence in systems with more than two electronic states.\cite{esch2020a}
 
At the end of each nuclear time step, we build the fully coherent electronic density matrix $\boldsymbol{\rho}^c(t+\Delta t)$ in the basis of adiabatic electronic states by taking the outer product of the current electronic wave function with itself.  Then, we form a target density matrix, $\boldsymbol{\rho}^d(t+\Delta t)$, with off-diagonal elements scaled by a scaling factor, $0 \leq w_{ij} \leq 1$, that incorporates the Gaussian decay of the coherence:
\begin{equation}
\rho_{ij}^{d}=w_{ij}(\tau_{ij},\dot{\rho}_{ii},\dot{\rho}_{jj}) \rho_{ij}^{c}(t+\Delta t).
\end{equation}
A complete definition of $w_{ij}(\tau_{ij},\dot{\rho}_{ii},\dot{\rho}_{jj})$ may be found in ref \citenum{tabgauss}.  The diagonal elements of the target density matrix are not scaled,
\begin{equation}
\rho_{ii}^{d}= \rho_{ii}^{c}(t+\Delta t).
\end{equation}

Having computed the target density matrix, $\boldsymbol{\rho}^d(t+\Delta t)$, we decompose this matrix as a sum of block-wise density matrices that represent pure superposition of subsets of the adiabatic states:
\begin{equation}
\label{eq:block}
\boldsymbol{\rho}^d=\sum_a P_a^{\mathrm{block}} \boldsymbol{\rho}_a^{\mathrm{block}}\,.
\end{equation}
Finally, we stochastically decide whether the wavefunction will collapse into one of pure states represented by these blocks. A random number between 0 and 1 is selected and compared to the probabilities, $\{P_a^{\mathrm{block}}\}$, to determine which state the electronic subsystem collapses into. 

After collapse, the entire momentum vector is scaled to ensure conservation of energy {with a factor of $\sqrt{1-\frac{E_{\mathrm{pot}}^{\mathrm{new}}-E_{\mathrm{pot}}^{\mathrm{old}}}{E_{\mathrm{kin}}}}$, adjusting the kinetic energy $E_{\mathrm{kin}}$ by the difference of TDCI electronic energies $E_{\mathrm{pot}}$ after and before the collapse.}
This block-wise algorithm allows TAB to correctly describe the decay of coherences between all pairs of states---even when many states are present, pairs of states with parallel PESs maintain coherence indefinitely, while very non-parallel pairs lose coherence rapidly.\cite{esch2020a} {Like the parent Ehrenfest method, a single TAB trajectory is not meaningful.  However, averaging over an ensemble yields an accurate treatment of decoherence, with subset of trajectories ultimately collapsing to individual states or sets of parallel states after a physically meaningful decoherence time.}

In this work, we use a fully {\em ab initio} method to propagate the electronic wave function: time-dependent configuration interaction.\cite{tdci} In TDCI, the wave function is expressed as a linear combination of Slater determinants, $\Phi_I$, with time-dependent expansion coefficients $C_I(t)$,
\begin{equation}
\psi(t) = \sum_I C_I(t) \Phi_I\,.
\end{equation}
We recast the Schr\"{o}dinger equation in the form
\begin{equation}
i\dot{\boldsymbol{C}}(t) = {\boldsymbol{H}_{\mathrm{e}}}(t){\bf{C}}(t)
\end{equation}
and solve the equation numerically using the symplectic split operator integrator.\cite{blanes2006symplectic} In this work, we select a complete active space (TD-CASCI) ansatz.  Our GPU-accelerate implementation is in the TeraChem electronic structure software package.\cite{terachem21,terachem_details,durden2022,fales2015gpu,tdci,Hohenstein2015} 

{To derive an expression for the forces experienced by the classical nuclei, we must note that the time-dependence of the electronic Hamiltonian matrix arises not only from an explicit dependence of the Hamiltonian operator on $\boldsymbol{R}(t)$, but also a dependence of the many-electron basis on the position-dependent orbitals, $\boldsymbol{X}(\boldsymbol{R}(t))$.  Then, starting from Eq. \ref{eq:eom} we can derive the expression for the $\mu$th element of the force,
\begin{equation}
F_{\mu} = -\boldsymbol{C}^\dagger \frac{\partial \boldsymbol{H}_{\mathrm{e}}}{\partial R_\mu} \boldsymbol{C} - \left( \boldsymbol{C}^\dagger \frac{\partial \boldsymbol{H}_{\mathrm{e}}}{\partial \boldsymbol{X}} \boldsymbol{C} \right) \frac{\partial \boldsymbol{X}}{\partial R_\mu}.
\end{equation} 
All terms are computed using existing time-independent gradient code, with only minor modifications.\cite{Hohenstein2015}}

 We have previously applied {\em ab initio} TAB to predict the ultrafast electronic diffraction spectrum of the photodissociation of cyclobutanone\cite{cycbut_ours} in the context of a community prediction challenge.\cite{miao2024casscf,martin2024photofragmentation} Our prediction, though slightly too fast, is in good agreement with experiment.\footnote{Preliminary experimental data communicated by Thomas Wolf.} However, that problem involved a small number of electronic states and only short-lived coherences. The present work represents the first application of TAB to describe long-lived electronic coherences in a system with many electronic states, the class of problems for which it is almost uniquely well suited.

\subsection{Thiophene's electronic structure}

Despite its apparent simplicity, previous theoretical studies reveal that accurately describing the electronic excited states of thiophene (C$_4$H$_4$S) can be quite challenging. The HOMO ($\pi_1$) and HOMO-1 ($\pi_2$) orbitals, from which the low-lying excitations originate, have $\pi$ character. The excitations lead not only to antibonding valence orbitals ($\pi^*$ and $\sigma^*$), but also to a set of low-lying Rydberg orbitals. This puts additional demands on the atomic basis set and wave function expansion. The difference in electronic correlation energy between the valence and Rydberg states is quite large, as demonstrated in many past studies based on configuration interaction\cite{PALMER1999275}, complete active space self-consistent field (CASSCF)\cite{SERRANOANDRES1993125,heller2021exploring,pastore2007multireference}, multireference perturbation theory \cite{kaufman_multiphoton, parkes2024simple}, coupled cluster theory \cite{pastore2007multireference, holland2014excited}, and other methods.\cite{thioioniz, salzmann2008excited, kleinschmidt2002spin, jones2019electronic} 

\begin{figure}
\includegraphics[width=0.5\textwidth]{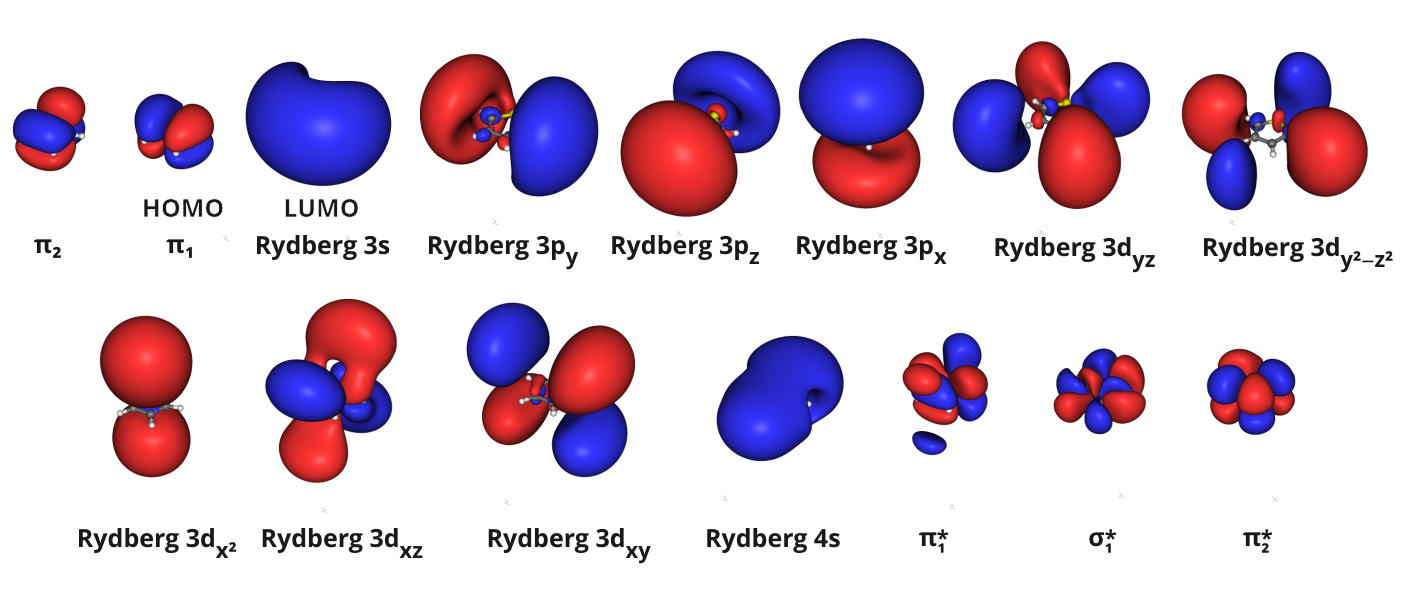}%
\caption{Molecular orbitals calculated by the FOMO-CASCI method at the S$_0$ minimum geometry.}%
\label{fig:orbs}
\end{figure}

For computational feasibility, we must choose a relatively inexpensive electronic structure method for use in conjunction with TAB. 
To this end, we choose the time-dependent (TD-) floating occupation molecular orbitals (FOMO-) complete active space configuration interaction (CASCI) approach.\cite{granucci2001fomo,fomo}  We use an active space of 4 electrons in 15 orbitals.  The space contains the full (four orbital) $\pi$ space, a full set of $n=3$ Rydberg orbitals (s, p, and d), a $\sigma^*$ orbital, and a 4s Rydberg orbital.  Figure \ref{fig:orbs} depicts the active space molecular orbitals (MOs) calculated at the S$_0$ minimum. (We define the C$_{2}$ axis of rotation is the z-axis, and the molecular ring is in the yz plane.) 
We use the cc-pVDZ basis, extended with Rydberg basis functions positioned at the molecular center of mass (COM). These were obtained by a contraction of a set of 8s8p8d diffuse functions\cite{kaufmann1989universal} with exponents optimized using the methodology of Roos et al.\cite{roos1996multiconfigurational} in the OpenMolcas software package.\cite{openmolcas23}  The FOMO temperature parameter is set to 0.15 a.u.  This value was empirically chosen because the resulting orbitals produced excitation energies in reasonable agreement with experiment, as detailed below.

To analyze the accuracy of the FOMO-CASCI excitation energies, we compare the calculated vertical excitation energies with the high-resolution absorption spectrum from Holland et al.\cite{holland2014excited} in Table \ref{tab:ene}. (Transition dipole moments are given in Table S1.)  The absorption spectrum contains two broad peaks centered at 5.5 and 7.0 eV. The first one is dominantly composed of valence-valence transitions (to $\pi_1^*$ and $\sigma^*$ states {-- lower part of Table \ref{tab:ene})}, with the ${\pi_1 \xrightarrow{} \mathrm{3s}}$ transition on its high-energy tail. The second one then arises from a wide range of Rydberg excitations and a pair of $\pi$ $\xrightarrow[]{}$ $\pi_2^*$ states. Looking at the FOMO-CASCI results, the method faithfully captures the Rydberg excitations to within a few tenths of an eV. However, FOMO-CASCI predicts the valence transitions to be ~2 eV higher than observed experimentally. This is due to profound dynamic electron correlation effects which are observed mainly for the valence states, a common and well-known error in CASSCF and CASCI results, discussed above. Despite this error in excitation energy, the MO shapes, relative transition dipole moments, and shapes of the PESs are expected to be captured adequately. We will address the impact of the overestimation of the valence states on dynamical results below.

\begin{table}[h]
\centering
\begin{tabular}{llcc}
\toprule[1.5pt]
& &  \multicolumn{2}{c}{Energy (eV)}\\
State & Character & FOMO-CASCI  & Exp.\cite{holland2014excited} \\
\midrule[0.75pt]
& Rydbergs &&\\
S$_1$ & $\pi_1 \xrightarrow{}$ 3s & 5.73 &  5.99 \\ 
S$_2$ & $\pi_1 \xrightarrow{}$ 3p$_{y}$ & 6.06 & 6.71\\ 
S$_3$ & $\pi_2 \xrightarrow{}$ 3s & 6.26  &6.47\\ 
S$_4$ & $\pi_1 \xrightarrow{}$ 3p$_{x}$ & 6.33  & 6.60 \\ 
S$_5$ & $\pi_1 \xrightarrow{}$ 3p$_{z}$ & 6.33  &  \\ 
S$_6$ & $\pi_2 \xrightarrow{}$ 3p$_{y}$ & 6.58  &  \\ 
S$_7$ & $\pi_1 \xrightarrow{}$ 3d$_{yz}$ & 6.65 &   (7.67) \\ 
S$_8$ & $\pi_2 \xrightarrow{}$ 3p$_{x}$ & 6.76 &  6.87$-$7.12\\ 
S$_9$ & $\pi_2 \xrightarrow{}$ 3p$_{z}$ & 6.76 & 6.87$-$7.12 \\ 
S$_{10}$ & $\pi_1 \xrightarrow{}$ 3d$_{x^2}$/3d$_{y^2-z^2}$ & 6.78 & (7.24) \\ 
S$_{11}$ & $\pi_1 \xrightarrow{}$ 3d$_{y^2-z^2}$/3d$_{x^2}$ & 6.83 & (7.28) \\ 
S$_{12}$ & $\pi_1 \xrightarrow{}$ 3d$_{xy}$ & 6.95 &  7.4$-$7.7 \\ 
S$_{13}$ & $\pi_1 \xrightarrow{}$ 3d$_{xz}$ & 6.98 & 7.4$-$7.7\\ 
S$_{14}$ & $\pi_2 \xrightarrow{}$ 3d$_{yz}$ & 7.13 &   \\ 
S$_{15}$ & $\pi_2 \xrightarrow{}$ 3d$_{xz}$/$\pi_1^*$/$\pi_2^*$ & 7.14  &  \\ 
S$_{16}$ & $\pi_1 \xrightarrow{}$ 4s & 7.17 &  \\ 
S$_{17}$ & $\pi_2 \xrightarrow{}$ 3d$_{y^2-z^2}$/3d$_{x^2}$ & 7.28 &   \\ 
S$_{18}$ & $\pi_2 \xrightarrow{}$ 3d$_{x^2}$/3d$_{y^2-z^2}$ & 7.31 &   \\ 
S$_{19}$ & $\pi_2 \xrightarrow{}$ 3d$_{xy}$ & 7.46 &   \\ 
S$_{20}$ & $\pi_2 \xrightarrow{}$ 3d$_{xz}$/$\pi_1^*$ & 7.49 & \\ 
S$_{21}$ & $\pi_2 \xrightarrow{}$ 4s & 7.65 &  \\ 
\midrule[0.75pt]
& Valence states &&\\
S$_{22}$ & $\pi_1 \xrightarrow{} \pi_1^*$ & 7.70 &5.16  \\ 
S$_{23}$ & $\pi_1 \xrightarrow{} \sigma_1^*$ & 7.86 &  6.0-6.2 \\ 
S$_{24}$ & $\pi_2 \xrightarrow{} \sigma_1^*$ & 8.03 &6.0-6.2 \\ 

\bottomrule[1pt]
\end{tabular}
\caption{Vertical excitation energies of thiophene calculated at the ground-state optimized geometry using FOMO-CASCI. Experimental results correspond to 0-0 transition energies or band ranges.  Parenthesized energies correspond to theoretical best estimates from the same reference.} 
\label{tab:ene}
\end{table}

\section{Results and Discussion}
\subsection{Multiphoton excitation}
To generate the initial superposition of states for subsequent TAB dynamical simulations, we explicitly model the multiphoton excitation of thiophene from its electronic ground state by the intense {pump} pulse employed in the experiment.  
{Our simulation of the pump is performed with frozen nuclei.}
{We use TD-FOMO-CASCI to describe the many-electron dynamics}, with the pulse modeled in the electric dipole approximation by adding an explicit time-dependent electric field to the Hamiltonian. Explicitly modeling the excitation process in this way implicitly includes many of the intricacies of strong-field multiphoton excitation: the bandwidth of the pulse (see fig. S1), the selection rules for multiphoton processes, any resonant enhancements, and the AC Stark effect.  Here we have used a transform-limited pulse with parameters corresponding directly to the experiment: a frequency of 1.55 eV, field intensity of 1.5E+13 W/cm$^2$, and pulse duration of 10 fs.  Each of the 50 initial geometries (along with momenta to be used in the subsequent TAB simulations) were sampled from the zero-temperature vibrational Wigner distribution function, computed in the harmonic approximation based on frequencies calculated at the MP2 level. The direction of the linearly polarized pulse was randomized. The nuclei were frozen during our 40 fs long simulations {to quantify the electronic response of each initial geometry to the field.} The simulations were started 15 fs before the pulse is to reach maximum intensity. At the end of the simulations, any remaining ground state character was projected out of the CI wavefunctions, and the {renormalized} superpositions were used as initial conditions for the subsequent TAB molecular dynamics simulations. 

\begin{figure}[h]
\centering
\includegraphics[width=0.45\textwidth]{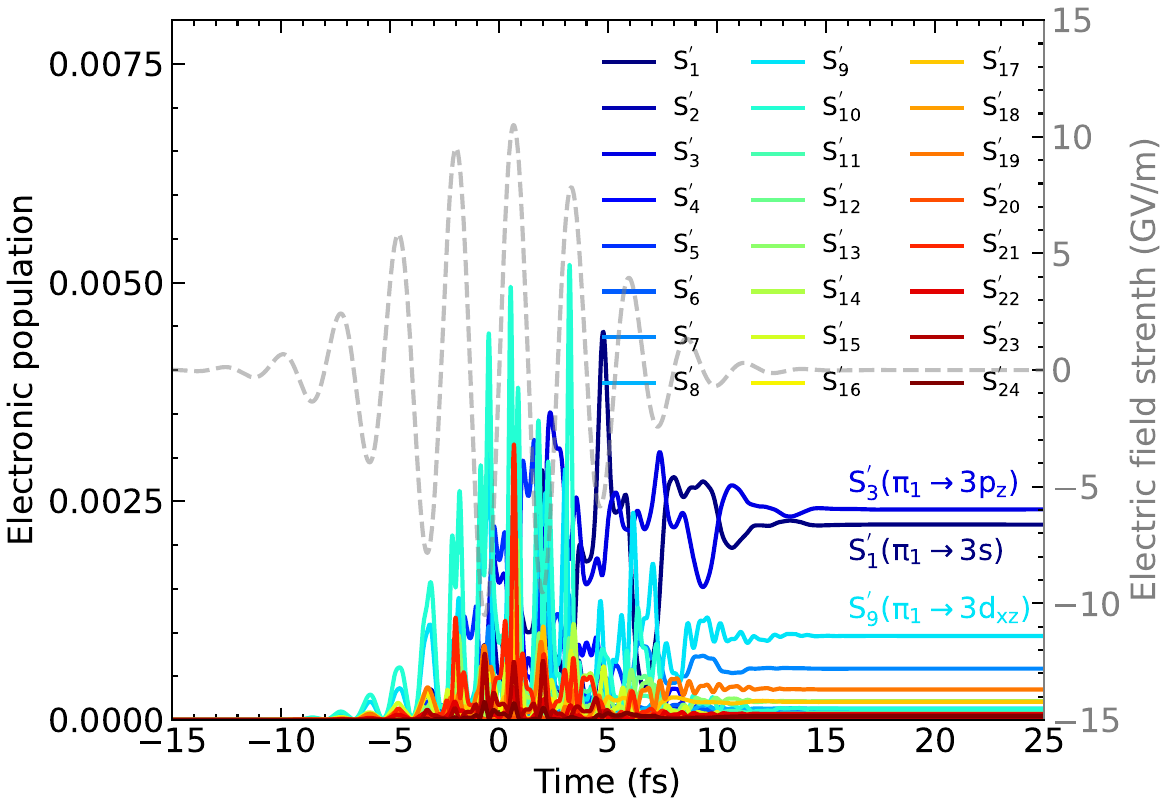}
\caption{The populations of electronic excited states plotted as a function of time from a selected TDCI simulation of excitation by the pump pulse. The field is shown as a dashed gray line. Selected state characters are printed in parentheses.}%
\label{fig:extraj2}
\end{figure}

State populations as a function of time are shown for a {single} selected representative of such a simulation in Fig.\ref{fig:extraj2}. Oscillations in state populations are observed to peak with the field intensity. After the field dies down, one is left with a superposition of excited states and any remaining ground state character. The final state populations depend not only on the molecular geometry but also on the polarization direction of the laser (See figs. S2-S4). In the present case, as in many of the trajectories, the population primarily settles in Rydberg states.  The populations of high-lying valence states oscillate with the laser field as well (e.g., S$_{21}^{\textnormal{\textquotesingle}}$, $\pi_2 \xrightarrow{} \pi_1^*$ in red), but these oscillations do not lead to significant population in the valence states at the end of the simulations. 

\begin{figure}[h]
\centering
\includegraphics[width=0.45\textwidth]{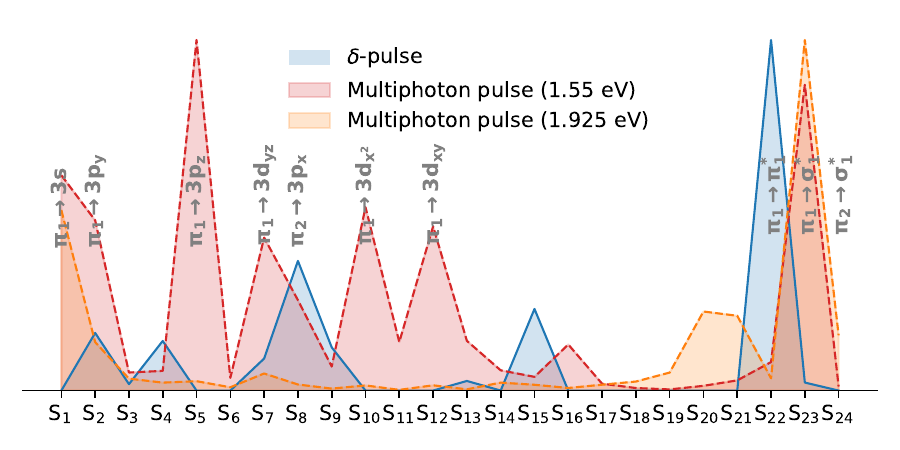}
\caption{The transition probabilities computed for a $\delta$-pulse (i.e., S$_0$$ \xrightarrow{}$S$_n$ oscillator strengths, blue) and two transform-limited pulses (plotting spherically-averaged final S$_n$ populations from TDCI simulations) at 1.55 eV (used in the experiment and simulations, red) and 1.925~eV (tuned to allow $4h\nu$ excitation to the S$_{22}$ valence state, orange). These probabilities were computed at the ground state minimum geometry using the FOMO-CASCI  method, averaging over all polarization angles.} %
\label{fig:strengths}
\end{figure}

The dominant Rydberg character is not to be expected from the one-photon transition dipole moments, but becomes important for multiphoton excitation by our strong laser pulse.  This can be seen in Fig.\ref{fig:strengths}, which compares the one-photon oscillator strengths (determined by exciting via a $\delta-$function pulse) with spherically-averaged populations for multiphoton excitation of the ground-state minimum energy geometry. Upon multiphoton excitation by our intense 1.55 eV pulse, a wider range of Rydberg states is excited while the role of valence states is diminished compared to one-photon excitation by the $\delta-$pulse.  It is particularly noteworthy that excitation to S$_{22}(\pi_1 \xrightarrow{} \pi_1^*$) is not significant for our intense pulse.  Excitation to the valence S$_{23}$ state ($\pi_1 \xrightarrow{} \sigma_1^*$) is more signficant with our multiphoton pulse.  However, the subsequent TAB simulations indicate that population of this state results in ring-opening reactions, and therefore, it appears to be an unlikely candidate to be involved in a long-lived electronic coherence due to rapid loss of vibrational overlap. 

As noted above, the excitation energy of S$_{22} (\pi_1 \xrightarrow{} \pi_1^*$) state is overestimated at the FOMO-CASCI level of theory.  In fact, the energy is so high that in our simulations S$_{22}$ is only accessible by 5$h\nu$ excitation (rather than 4$h\nu$, as would be possible in the experiment).  One might hypothesize that 4$h\nu$ excitation might more efficiently populate the $\pi_1 \xrightarrow{} \pi_1^*$ state, if the energies were properly aligned.  To confirm the diminished role of the S$_{22}$ ($\pi_1 \xrightarrow{} \pi_1^*$) transition, we performed additional simulations using a pulse centered at 1.925 eV    , therefore targeting S$_{22}$ for 4h$\nu$ excitation. The results confirm low probability of excitation to S$_{22}$. For more details on initial superpositions and source data see Supporting Material. We also refer the reader to past theoretical work on thiophene, where the method of adiabatic elimination was used to simulate this process.\cite{kaufman_multiphoton} These results also indicate large changes in the absorption spectrum with the the photon order. 

\begin{figure}[h]
\centering
\includegraphics[width=0.49\textwidth]{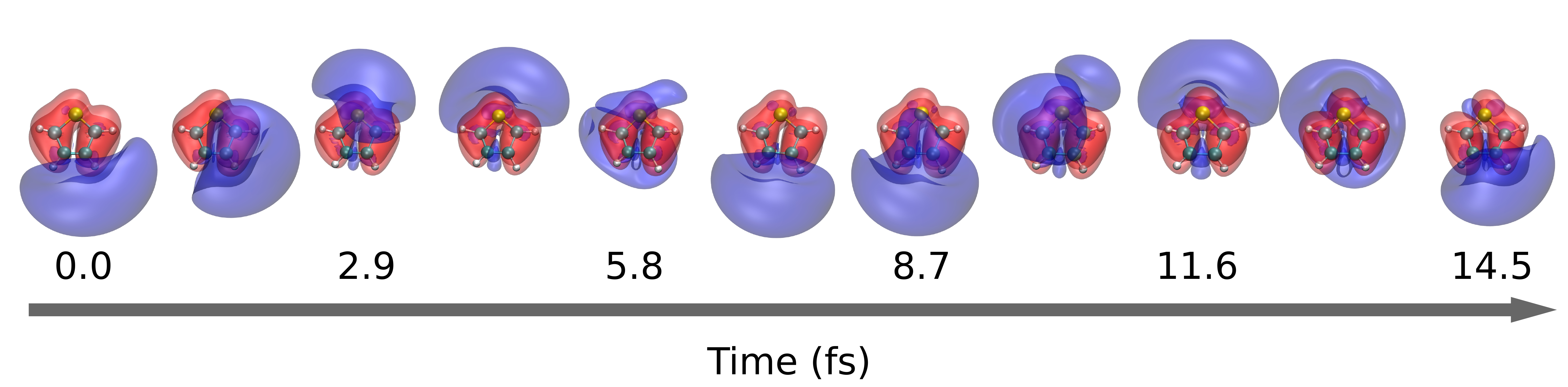}
\caption{Time evolution of the superposition formed in Fig.\ref{fig:extraj2} {during a single TAB trajectory.} We visualize positive (blue) and negative (red) electron density difference relative to the electronic ground state.}%
\label{fig:tab2}
\end{figure}

To gain better understanding of the resulting electronic coherence, we present a visualization in Fig.\ref{fig:tab2}. Specifically, we display the {TAB} time evolution of the electronic density difference between the excited and the ground state for {the selected initial superposition, which formation we have previously shown} in Fig.\ref{fig:extraj2}. It is predominantly a superposition of states S$_1$, S$_3$ and S$_9$, with Rydberg 3s, 3p$_z$, and 3d$_{xz}$ character, respectively. The snapshots cover 14.5 fs of TAB dynamics, which corresponds to two oscillation periods dictated by an energy gap of 0.6 eV between the S$_1$ and S$_3$ states. The coherent superposition results in a flow of electronic density (blue) from the upper side of the molecule to the lower one and back again. Additionally, interference with the 3d$_{xz}$ state introduces motion to appear in front and behind the molecular plane in between the two cycles. The negative part of the difference density (red) represents the hole from which the excited electron originates, i.e. the HOMO $\pi_1$ orbital. There is no noticeable evolution in the red part since both excitations share the same hole orbital. This is vital for experimental observation of coherences through ionization yield since a coherence would only be observable if the same final cationic state is accessible from all of the stationary states involved in the superposition.\cite{vacher_1}

\subsection{TAB dynamics}

\begin{figure}
\includegraphics[width=0.45\textwidth]{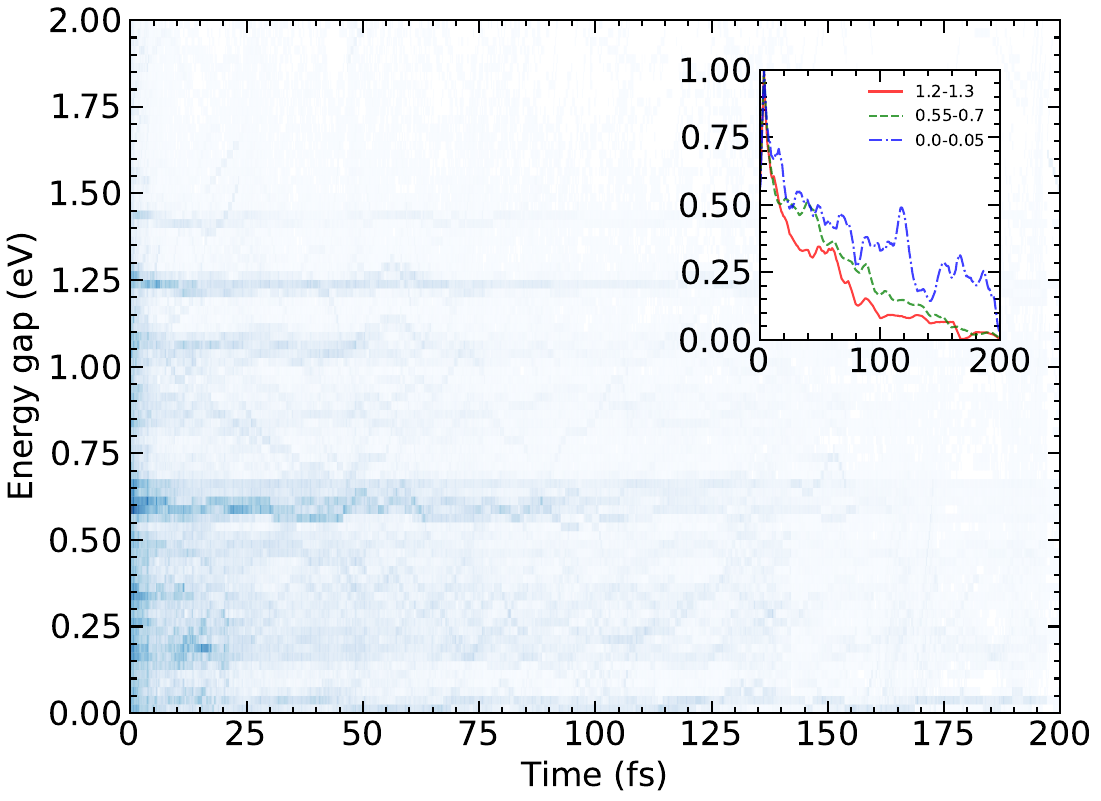}%
\caption{The ensemble-averaged absolute coherence (Eq. \ref{eq:blueness}, as a function of the time (x-axis) and pairwise energy gap (y-axis).  The inset plot shows the mean-averaged (5 fs) integrated total of the signal within the given energy ranges, each normalized to its maximal value.}%
\label{fig:coh}
\end{figure}

In order to capture the fully coupled nuclear-electronic dynamics of the whole ensemble of molecules, we simulated 50 trajectories for 200 fs via TAB using the initial conditions generated in the previous subsection. To show the electronic coherences we devised Fig. \ref{fig:coh} which displays pairwise energy gaps of concurrently populated electronic states over time. The "blue-ness" of each pixel corresponds to the ensemble-averaged absolute coherence as a function of both time and energy gap between states ($\Delta E$), defined as
\begin{equation}
\label{eq:blueness}
    b(t,\Delta E) = \sum_{ijI} \lvert a_i^I(t) a_j^{I*}(t) \rvert \delta(E_i-E_j-\Delta E) P_I,
\end{equation}
where $I$ indexes that 50 trajectories in the ensemble, $i$ and $j$ index electronic states, and $\delta$ is the Dirac delta function.
Additionally, we weigh each trajectory contribution by the amount of excited state population after the laser pulse excitation, $P_I$, reflecting higher probability of occurrence. After a trajectory has collapsed to a single state, we no longer plot its signal to remove experimentally unobservable coherences, like those created following passage through conical intersections. Long-lasting coherences are observed as horizontal lines.  Such coherences are observed at several gaps, the strongest of which can be seen at 0.6 eV.  Collapses, triggered by the non-parallelity of the states (Eq. \ref{eq:decohtime}) or changes in state populations, result in the decay of these lines on the timescales ranging from tens of fs up to~100 fs. {Note that collapses represent decoherence mechanism c), discussed in the intro---the loss of coherence is due to the decay of nuclear overlap between populations on different electronic states.} 
A video visualization of the coherent oscillations of the electronic wave packet and subsequent loss of coherence over time is included as supplementary material. 

Each line can be assigned to unique pairs of states, which exhibit a nearly constant energy gap throughout the simulations. This is one of the requirements of long-lasting coherences since any difference in potential surfaces leads to faster decoherence. This means that any signal on the coherence plot diverging from a straight line usually suddenly disappears, indicating collapse of the electronic wavefunction. Thiophene is a highly symmetric system with many parallel excited state potentials which can support the observed coherences. We can show this by plotting the PES scans along normal modes (see Fig S5). We can expect that phase-dependent ionization yield, which is ultimately measured by the experiment, would exhibit oscillations corresponding to these energy gaps.

By further analyzing the signals, we observe that \mbox{$\pi_1 \xrightarrow{}$ 3s} (S$_1$) excited state is usually involved in the gap pair. For instance, the strongest signal at 0.6 eV arises due to coherence between S$_1$ with S$_5$ ($\pi_1 \xrightarrow{}$ 3p$_z$) as its partner. The gap near 0.05 eV mainly represents a gap between two p or d Rydberg states, very close in energy. 
The inset plot additionally shows integrated signal strength within selected ranges, normalized to the maximal value. 
We fitted an exponential decay function to each curve, ignoring the first 10 fs of the dynamics to circumvent noisy data in this region. This provided average coherence half-life $\tau_{1/2}$ of around 50 fs for the band at 0.6 eV and 42 fs for the band at 1.25 eV. This is slightly smaller than the inferred experimental value of around 100 fs. 
Apart from apparent sources of simulation error (e.g., errors in state energies), it is worth noting that our approach cannot capture recoherence, which would result in longer-observed coherences. More generally, the loss of overlap in the nuclear wavefunction and later realignment cannot be captured with simple independent trajectory methods, such as TAB. Thus, investigating the role of recoherence will be a future research direction.

\subsection{Assignment of Coherences}

Finally, we turn our attention to assigning the coherences observed in the experiment of Kaufman et al.  In order to be consistent with the observed experimental data, several things must be true:  a) both states involved in the coherence must be accessible upon multi-photon excitation; b) the PESs of the two states must be sufficiently parallel to support a long-lived coherence; c) the real-life gap between the states must correspond to the experimentally observed energy gap, 1.5 eV;\cite{kaufman2023longthio} and d) for the coherence to be experimentally observable, the two states in question must correspond to the same cationic core.\cite{vacher_1}  

The above results strongly support the assignment of the experimentally observed coherences to two or more Rydberg state.  First, we can eliminate the valence $\pi_1\xrightarrow{}\pi_1^*$ state because it is not readily excited by 4 photons (Fig. \ref{fig:strengths}). The more accessible $\pi_1 \xrightarrow{} \sigma_1^*$ state can be eliminated from consideration because excitation to this state results in ring opening, which destroys nuclear wave function overlap, thus killing the coherence very rapidly. Analysis of our simulated products reveals that 90 \% of the molecules remain in the cyclical form at the end of the simulation while some of the molecules undergo ring-opening dissociation. This route has previously been studied in relation to the $\pi \xrightarrow{} \pi^*$ excitations which can undergo internal conversion to the $\pi \xrightarrow{} \sigma^*$ state, opening one of the C-S bonds.\cite{prlj2015excited,schnappinger2017ab} We also see involvement of these states in the ring-opening reaction.

\begin{table}[h]
\centering
\begin{tabular}{cclcc}
\toprule[1.5pt]
\multirow{2}{*}{$\Delta E$} &  \multirow{2}{*}{$\tau_{1/2}$(fs)}   & \multirow{2}{*}{Assigned states} & \multicolumn{2}{c}{Energy gap (eV)} \\ \cline{4-5} 
& & & FOMO-CASCI & Exp.\cite{holland2014excited} \\ 
\midrule[0.75pt]
\multirow{3}{*}{0.6} & \multirow{3}{*}{52} &  $\pi_1 \xrightarrow{}$ 3s \& $\pi_1 \xrightarrow{}$ 3p$_z$ & 0.60   &   0.61 \\
& &  $\pi_1 \xrightarrow{}$ 3p$_z$ \& $\pi_1 \xrightarrow{}$ 3d$_{xz}$ & 0.65  &  0.8--1.1    \\
& & $\pi_1 \xrightarrow{}$ 3p$_z$ \& $\pi_1 \xrightarrow{}$ 3d$_{xy}$ & 0.62 &  0.8--1.1\\
\hline
\multirow{2}{*}{1.25} &  \multirow{2}{*}{42} & $\pi_1 \xrightarrow{}$ 3s \& $\pi_1 \xrightarrow{}$ 3d$_{xz}$ & 1.25 &    1.4--1.7  \\
& & $\pi_1 \xrightarrow{}$ 3s \& $\pi_1 \xrightarrow{}$ 3d$_{xy}$  & 1.22 &  1.4--1.7 \\
\bottomrule[1pt]
\end{tabular}
\caption{Assignment of the most prominent long-lived coherences in Fig. \ref{fig:coh}. For each, the energy gap ($\Delta E$) and half-life of the observed decay ($\tau_{1/2}$) are provided. The signal half-lifes are obtained by fitting an exponential decay to the integrated signals shown in the inset of Fig.\ref{fig:coh}, neglecting the first 10 fs of the simulations. Within each band, several pairs of states that contribute to the coherence are listed by their state character, FOMO-CASCI energy gap (at the ground state minimum geometry), and corresponding experimental transition energy. The group of states with gap $\sim$1.25 eV plausibly corresponds to the observed experimental signal.\cite{kaufman2023longlivedelectroniccoherences} } 
\label{tab:asgn}
\end{table}

Table \ref{tab:asgn} summarizes the assignments of the Rydberg electronic states to the coherences observed in the simulation.  Also presented are the experimentally determined energy gaps corresponding to these pairs of states, determined from the previously discussed high-resolution absorption spectrum.
Considering all of the above factors, the observed 1.25 eV (FOMO-CASCI) gap between the 3s and 3d$_{xz}$ or 3d$_{xy}$ Rydberg states makes them good candidates for the source of the experimental signal. 

{Symmetry has been noted to play a role in the observation of coherences created by ultrafast passage through conical intersections.\cite{thio_neville2022formation}  To within sampling error, all spectroscopic selection rules are naturally incorporated into our simulation of the pump pulse, as are the effects of symmetry on the subsequent dynamics.  However, we have not explicitly computed the photoelectron yield. Thus, the degree to which these particular coherences modulate the photoelectron signal remains open for future work.}

\section{Conclusions}
Herein, a novel combination of TDCI simulations of multiphoton absorption and TAB simulations of the subsequent molecular dynamics reveals long-lived coherences in thiophene. Results show the sole involvement of Rydberg states, with coherence half-lives of around 50 fs, allowing us to directly assign the long-lived coherences observed in recent ultrafast experiments.\cite{kaufman2023longlivedelectroniccoherences} Though we observe coherences between many pairs of states, based on several considerations, the experimentally observed signal can be assigned to coherence between 3s and 3d Rydberg states. Here coherences between pairs of Rydberg states are long-lived because they share a common cationic core, and therefore nearly parallel PESs. The Rydberg electron itself orbits far enough from the nuclei that it exerts a relatively small force. {Our simulations still underestimate the experimentally observed half-lives, which might be due to lack of recoherence. We would like to address this issue in the future by some type of collective post-processing of trajectories.\cite{barbatti_s2024quantum}} It is worth noting that molecules in cavities may also achieve long-lived coherences, owing to dilution of the electron-phonon coupling in the resulting polariton state.\cite{Hu2022} The ability of the TAB approach to efficiently model dynamics on many electronic states and to discern between long- and short-lived coherences opens a range of opportunities in fields ranging from attoscience to quantum materials to photobiology.

\section*{Acknowledgments}
The authors acknowledge many valuable conversations with Tom Weinacht, Brian Kaufman, Ari Pereira, and Sam Vasquez. The authors gratefully acknowledge support from the National Science Foundation under grant NSF-1954519 and the Institute for Advanced Computational Science (IACS).  J.S. acknowledges a postdoctoral fellowship from IACS.

\section*{Supplementary Material}
See the supplementary material for following: details on multiphoton excitation calculations (pulse spectrum, initial superpositions, spherical sampling for different pulse energies), excited states properties, excited PESs along normal modes, basis set details, a video visualization of the coherent oscillations

\section*{Author Declarations}
\subsection*{Conflict of Interest}
The authors have no conflicts to disclose.
\subsection*{Author Contributions}
Jiří Suchan: Investigation (lead); Software (lead); Writing – original draft (equal); Writing – review \& editing (equal). Benjamin G. Levine: Conceptualization (lead); Writing – original draft (equal); Writing – review \& editing (equal).

\section*{Data Availability Statement}
The data that support the findings of this study are available from the corresponding author upon reasonable request.

\bibliography{citations}
\end{document}